\documentstyle[aps,prb,epsf,floats]{revtex}
\begin{document}
\twocolumn[\hsize\textwidth\columnwidth\hsize\csname@twocolumnfalse%
\endcsname
\title{Charge order, superconductivity, and a global phase
diagram of doped antiferromagnets}
\author{Matthias Vojta and Subir Sachdev}
\address{Department of Physics, Yale University\\
P.O. Box 208120, New Haven, CT 06520-8120, USA}
\date{November 8, 1999}
\maketitle
\begin{abstract}
We investigate the interplay between lattice-symmetry breaking and
superconducting order in a two-dimensional model of doped
antiferromagnets, with long-range Coulomb interactions and ${\rm Sp} (2N)$
spin symmetry, in the large-$N$ limit. Our results motivate the outline of a
global phase diagram for the cuprate superconductors. We describe
the quantum transitions between the phases, the evolution of
their fermion excitation spectrum, and the experimental
implications.
\end{abstract}
\pacs{PACS numbers:}
]
A number of recent experiments have found a rich variety of phases in the
cuprate superconductors~\cite{jtran,birg,imai}. The various ground
states can be distinguished by the manner in which they preserve,
or spontaneously break, three distinct and familiar symmetries of
the Hamiltonian: (a) the electromagnetic $U(1)$ symmetry, ${\cal
S}$, which is broken in the $d$-wave superconducting phase, but is
preserved in an insulating ground state; (b) the ${\rm SU}(2)$
spin rotation symmetry, ${\cal M}$, which is broken in
magnetically ordered phases; and (c) the symmetry of square
lattice translations and rotations, ${\cal C}$, which we will
consider broken if an observable invariant under ${\cal S}$ and
${\cal M}$, like the charge density, is not identical on every
site and every bond. We shall take the point of view here that all
the phases are conventionally characterized by the manner in which
${\cal S}$, ${\cal C}$, and ${\cal M}$ are broken, and have no
`exotic' properties or excitations, {\em i.e.\/}, in principle, an
appropriate electron Hartree-Fock/RPA/BCS theory, with
perturbative corrections, can be found; the anomalous finite
temperature ($T$) properties are then believed to be signatures of
quantum-critical points separating these
phases~\cite{SY,castro,zhang,laugh}.

This paper will describe the $T=0$, global phase diagram of
two-dimensional, doped antiferromagnets by discussing the
competition between phases in which one or more of the ${\cal S}$,
${\cal C}$, and ${\cal M}$ symmetries may be broken. Among our
results will be the complete quantitative solution of a
microscopic model of a doped antiferromagnet for the case where
the ${\cal M}$ symmetry is generalized~\cite{rs1} from ${\rm
SU}(2)$ to ${\rm Sp}(2N)$, (note ${\rm SU}(2) \cong {\rm Sp}(2)$)
and the large-$N$ limit is taken under a particular representation
of ${\rm Sp}(N)$. The simplifying feature of this limit is that it
restricts attention to the portion of the phase diagram (see
Fig.~\ref{fig1} below) where the ${\cal M}$ symmetry remains
unbroken; however, it does allow a realistic description of the
subtle and complicated interplay between the ${\cal C}$ and ${\cal
S}$ symmetries. Our results include ({\em i\/}) computation of the
doping dependence of the charge-ordering configuration and the
evolution of the ordering wavevector, ({\em ii\/}) computation of
the single-particle fermion spectrum, measurable in photoemission
experiments, in phases with ${\cal C}$ and ${\cal S}$ broken, and
({\em iii}) proposal of a quantum-critical field-theoretic model
to explain the recently observed~\cite{valla} anomalous $T$ and
frequency dependence of the photoemission line-width.

We will consider the following extended ``$t-J$'' Hamiltonian for
fermions, $c_{i \alpha}$, on the sites, $i$, of a square
lattice with spin $\alpha=1 \ldots 2N$
($N=1$ is the physical value):
\begin{eqnarray}
{\cal H} = \sum_{i > j} && \left[ -\frac{t_{ij}}{N} c_{i \alpha}^{\dagger} c_{j
\alpha} + {\rm H.c.}  + \frac{V_{ij}}{N} n_i n_j \right. \nonumber \\
&&~~~~~~~~+\left. \frac{J_{ij}}{N} \left( {\bf S}_i \cdot {\bf S}_j -
\frac{n_i n_j}{4N} \right) \right].
\label{e1}
\end{eqnarray}
Here $n_i = c_{i \alpha}^{\dagger} c_{i \alpha}$ is the on-site
charge density, and the spin operators ${\bf S}_i$ are fermion
bilinears times the traceless generators of ${\rm Sp} (2N)$. We
will be primarily concerned with the case where the fermion
hopping, $t_{ij}$, and exchange, $J_{ij}$, act only when $i,j$ are
nearest neighbors, in which case $t_{ij} = t$ and $J_{ij} = J$;
however, we will occasionally refer to cases with second neighbor
hopping ($t'$) or exchange ($J'$). The Coulomb interaction between
the electrons is represented by the on-site constraint $n_i \leq
N$, and the off-site repulsive interactions $V_{ij}$ which fall
off as the inverse separation between the sites. The $V_{ij}$ are
included to counter-act the phase separation tendency of the
$t$-$J$ model~\cite{ekl,sr,hm}, and play a key role in our
analysis. We shall be interested in describing the ground state of
${\cal H}$ as a function of its couplings and the average doping
concentration, $\delta$, which is fixed by $(1/N_s) \sum_i \langle
n_i \rangle = N(1-\delta)$, where $N_s$ is the (infinite) number
of sites.

The proposed phase diagram of ${\cal H}$ is shown in
Fig.~\ref{fig1}.
\begin{figure}
\epsfxsize=3.4in
\centerline{\epsffile{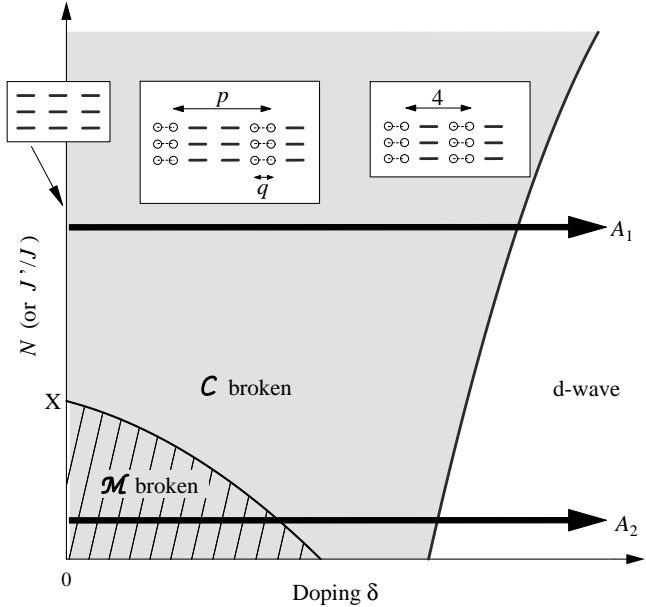}}
\caption{Schematic, proposed, ground state phase diagram of ${\cal H}$
as a function of the doping $\delta$ for physically reasonable values of
$t$, $J$ and $V$. The vertical axis represents a parameter
which measures the strength of
quantum spin fluctuations---it increases linearly with $N$ but can also be tuned
continuously by $J'/J$. The magnetic ${\cal M}$ symmetry is broken in the
hatched region, while ${\cal C}$ symmetry is broken
(with accompanying charge-density modulation)
in the shaded region; there are numerous additional phase transitions at which
the detailed nature of the ${\cal M}$ or ${\cal C}$ symmetry breaking changes - these
are not shown. For $\delta=0$, ${\cal M}$ symmetry is broken
only below the critical point $X$, while ${\cal C}$ symmetry is broken only
above $X$. The superconducting ${\cal S}$ symmetry
is broken for all $\delta > 0$ at large $N$;
for smaller $N$, the ${\cal S}$ can be restored at small $\delta$
by additional ${\cal C}$ breaking along the vertical axis for the
states in the inset--this is not shown.
The superconductivity is pure $d$-wave only in the large $\delta$ region were
${\cal C}$ and ${\cal M}$ are not broken. The arrow $A_1$ represents the path
along which quantitative results are obtained in this paper, while $A_2$
is the experimental path. The nature of the ${\cal C}$ symmetry breaking along path
$A_1$ is also sketched: the thick and dashed lines indicate varying
values of $|Q_{ij}|$ (proportional to the bond charge density) on
the links, while the circles represent $b_i^2$ (proportional to
the site hole density).
The charge densities on the links and sites not shown
take values consistent with the symmetries of
the figures shown. We expect that the nature of the ${\cal C}$
symmetry breaking will not change significantly as we move from
$A_1$ to $A_2$, and across the phase boundary where ${\cal M}$
is broken: this suggests the appearance of collinearly polarized
spin-density waves, which break both ${\cal C}$ and ${\cal M}$,
and which undergo an
`anti-phase' shift across the hole-rich stripes~\protect\cite{jan}.
}
\label{fig1}
\end{figure}
First, consider the vertical line, $\delta = 0$. Below $X$,
magnetic N\'{e}el order is present and so ${\cal M}$ is broken;
however, the charge densities are identical on every bond and
site, and so ${\cal C}$ is preserved, as is ${\cal S}$ because the
ground state is an insulator. Above $X$, there is a transition to
a quantum paramagnet and ${\cal M}$ symmetry is restored; this
transition was studied in Refs.~\onlinecite{rs0,rs1}, and it was
argued that ${\cal C}$ was necessarily broken in the quantum
paramagnet leading to spin-Peierls order. We can also view the
spin-Peierls order as a {\em bond-centered} charge-density wave
with a $2 \times 1$ unit cell~\cite{fn1}. Recent work~\cite{kotov}
has shown strong evidence for this order in the $N=1$ model with
$J' > 0 $.

We now describe the evolution of the ground state with increasing
$\delta$ along $A_1$. The large-$N$ limit is taken, as described
earlier~\cite{sr}, by minimizing the saddle point free energy with
respect to the site charge density $N(1-b_i^2) = \langle n_i
\rangle$ and the complex bond pairing amplitude $N Q_{ij} =
\langle {\cal J}^{\alpha\beta} c_{i\alpha}^{\dagger} c_{j
\beta}^{\dagger} \rangle/(b_i b_j)$ (where $b_i^2$ is the hole
density at site $i$ and ${\cal J}$ denotes the ${\rm Sp
}(2N)$-invariant antisymmetric tensor), while maintaining certain
local and global constraints. There have been a  number of related
earlier mean-field studies~\cite{bza}, but they have all (with the
exception of Ref.~\onlinecite{sr}) restricted attention to the
case where $b_i$ and $|Q_{ij}|$ are spatially uniform (note that
$|Q_{ij}|$ has the same symmetry signature as the bond charge
density, and is therefore a measure of its value). However such
solutions are usually unstable, and at best metastable, at low
doping; here we have attempted to find the true global minima of
the saddle-point equations, while allowing for arbitrary spatial
dependence: such a procedure leads to considerable physical
insight, and also leads to solutions in accord with recent
experimental observations.

\begin{figure}
\epsfxsize=2.8in
\centerline{\epsffile{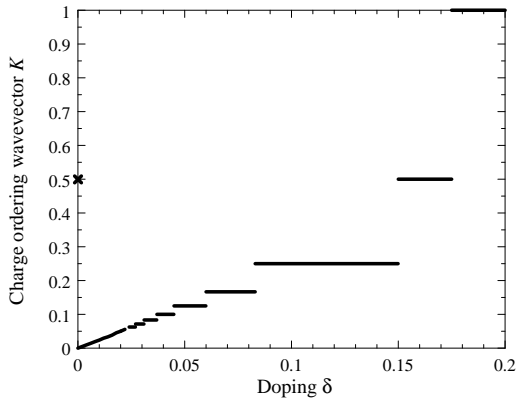}}
\caption{
The charge-ordering wavevector, $K$, (in reciprocal lattice
units) as a function of $\delta$ at $N=\infty$ for
$t/J = 1.25$, $V_{nn}/t = 0.6$ (where $q=2$).
For the states in Fig.~\protect\ref{fig1}, $K=1/p$.
We have $K=1/2$ at $\delta =0$.
The $K=1$ value at large $\delta$ has ${\cal C}$
symmetry restored, and is a pure $d$-wave superconductor.
For other values of parameters, the $K=1/2$ plateau does not occur, and there
is a direct jump from $K=1$ to $K=1/4$ (or smaller)
(see Fig.~\protect\ref{fig3}).
}
\label{fig2}
\end{figure}

First, at $\delta =0$ along $A_1$ we find the fully dimerized,
insulating spin-Peierls (or $2 \times 1$ bond charge-density wave)
solution~\cite{rsnp} in which $|Q_{ij}|$ is non-zero only on the
bonds shown in Fig.~\ref{fig1}. Moving to small non-zero $\delta$
along $A_1$, our numerical search always yielded lowest energy
states with ${\cal C}$ broken, consisting of {\em bond-centered
charge-density waves}~\cite{ws} with a $p \times 1$ unit cell, as
shown in Fig.~\ref{fig1}. We always found $p$ to be an even
integer, reflecting the dimerization tendency of the $\delta=0$
solution. Within each $p \times 1$ unit cell, we find that the
holes are concentrated on a $q \times 1$ region, with a total
linear hole density of $\rho_{\ell}$. A key property is that $q$
and $\rho_{\ell}$ remain finite, while $p \rightarrow \infty$, as
$\delta \rightarrow 0$. Indeed, the values of $q$ and
$\rho_{\ell}$ are determined primarily by $t$, $J$, and the
nearest-neighbor value of $V_{ij} = V_{nn}$, and are insensitive
to $\delta$ and longer range parts of $V_{ij}$. For $\delta
\rightarrow 0$, we found that $q=2$ was optimum for a wide range
of parameter values, while larger values of $q$ ($q \ge 4$) appear
for smaller values of $V_{nn}$; specifically we had $q=2$,
$\rho_{\ell} = 0.42$ at $t/J = 1.25$, $V_{nn}/t  = 0.6$, and $q=4$
, $\rho_{\ell} = 0.8$ at $t/J = 1.25$, $V_{nn}/t  = 0.5$. The
limit $V_{nn} \rightarrow 0$ leads to $q\rightarrow \infty$ which
reflects the tendency to phase separation in the ``bare'' $t-J$
model. The evolution of $p$ with $\delta$ is shown in
Fig.~\ref{fig2}.
Note that there is a large plateau at $p=4$ around doping $\delta = 1/8$,
and, for some
parameter regimes, this is the last state before ${\cal C}$ is
restored at large $\delta$; indeed $p=4$ is the smallest value of $p$
for which our mean-field theory has solutions with $b_i$ not
spatially uniform. Experimentally~\cite{jtran,birg}, a pinning of the charge
order at a wavevector $K=1/4$ is observed, and we
consider it significant that this value emerges naturally from
our theory.

Our large-$N$ theory only found states in which the ordering wavevector
$K$ was quantized at the rational plateaus in Fig.~\ref{fig2}. However, for
smaller $N$ we expect that irrational, incommensurate, values of $K$ will
appear, and interpolate smoothly between the plateau regions.

In our large-$N$ theory, each $q$-width stripe above is a
one-dimensional superconductor, while the intervening
$(q-p)$-width regions are insulating. However, fluctuation
corrections will couple with superconducting regions, yielding
an effective theory discussed in Section VII of
Ref.~\onlinecite{mpaf} with their dimensionless parameter $K \sim
N$. This implies that Josephson pair tunneling between the one-dimensional
superconductors is a relevant perturbation at sufficiently large $N$, leading to
two-dimensional superconductivity.
However, the bare pair-tunneling amplitude is exponentially small in
$p$, while the Coulomb interaction between the hole-rich regions
falls off only as $1/p$---the latter can then dominate for smaller
$N$ and $\delta$, leading to further ${\cal C}$ breaking along the
vertical stripe directions, and a transition to a two-dimensional
insulating state with ${\cal S}$
restored and an even number of electrons per unit cell.
Such an insulating state is more likely at rational
$\delta$, when the charge-ordering period along the
vertical stripe direction is commensurate with the
lattice.

\begin{figure}
\epsfxsize=3.3in
\centerline{\epsffile{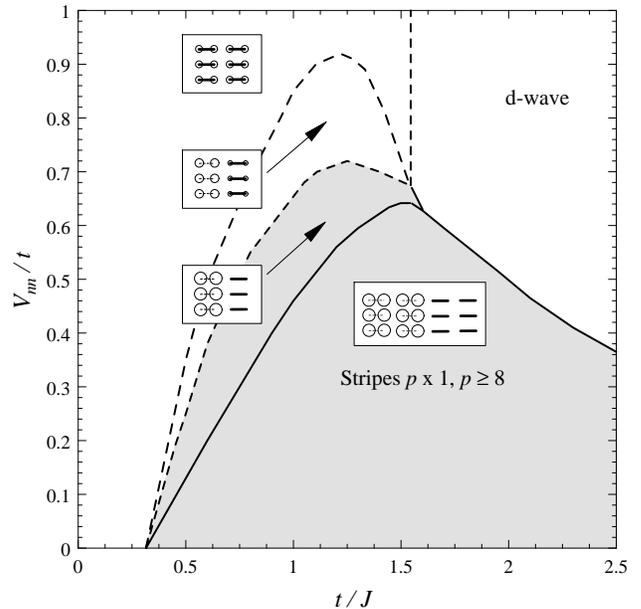}}
\caption{Ground states of ${\cal H}$ at $\delta = 1/8$
and $N=\infty$. Full (dashed)
lines indicate first- (second-) order transitions.
All states have
superconducting order, but the superconductivity is
one-dimensional (only at $N=\infty$) in the phases
with full stripe order (shaded).
}
\label{fig3}
\end{figure}

We show a fixed $\delta =1/8$, large $N$, cross-section of our results in
Fig.~\ref{fig3}.
The transition from a $d$-wave superconductor, with ${\cal C}$ unbroken,
to the fully-formed $p \times 1$ stripes discussed above can
either be first-order, or via intermediate states with partial
stripe order. In the latter case, there is first a continuous
transition to a state with ${\cal C}$ symmetry breaking at $p=2$ -- every
site is equivalent in such a state, and so the site charge density
is uniform while there is a modulation in the bond charge density;
this state can also be viewed as possessing coexisting
superconducting and spin-Peierls order~\cite{sr}. To our knowledge
a $p=2$ charge-ordered superconducting state has not been
experimentally detected, but a search for one should be
worthwhile. There is a second second-order transition to
$p=4$ state with partial stripe order, before the fully-formed
$p=4$, $q=2$ state with intervening insulating stripes appears
(Fig.~\ref{fig3}).
Larger values of $V$ suppress phases with a non-uniform distribution of site
charge densities; such phases also disappear in the limits of small $t/J$, and
$t/J\rightarrow\infty$.

\begin{figure}
\epsfxsize=3.3in
\centerline{\epsffile{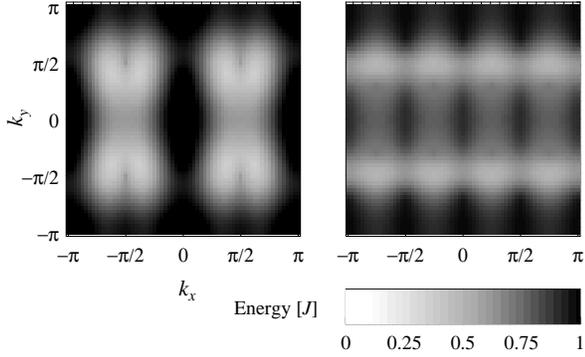}}
\caption{
Dispersion of the fermionic excitation spectrum obtained from
the mean-field solution of ${\cal H}$ at $\delta = 1/8$.
Left: $p=2$ phase at $t/J=1.5$, $V/t=1$, i.e., very close to the
transition to the $d$-wave phase.
Right: $p=4$ with partial stripe order at $t/J=1.25$, $V/t=0.8$.
Both spectra are fully gapped. Here, $k_x$ is the momentum parallel to
the charge-ordering wavevector ({\it i.e.}, the stripes run in $y$
direction).
}
\label{fig4}
\end{figure}

We now discuss the fermion excitation spectrum
in the states found above. The $d$-wave superconductor of course
has gapless, linearly dispersing fermion excitations along the $(1,\pm 1)$
directions in the Brillouin zone.
The various charge-ordered phases in general show a gapped spectrum,
see Fig.~\ref{fig4}.
In the fully striped phases the fermion energy is independent of
$k_x$ (the momentum perpendicular to the stripes), the dispersion
minimum is near $(0,\pm 1/4)$.
In the $p=2$ phases (and also for $p=4$ and partial stripe
order) the minimum of the energy is at complex values of
pairing amplitudes $Q_{ij}$; these states break time-reversal
symmetry $\cal T$ and their fermionic excitations are fully gapped.
However, if we restrict our attention to states without $\cal
T$-breaking, then upon decreasing $t/J$ in the $d$-wave superconductor
(at large $V$, see Fig.~\ref{fig3}) the gapless fermions survive across the
$\cal C$-breaking transition to the $p=2$ phase;
the excitation gap then opens at smaller $t/J$ (i.e. at a finite dimerization).

Finally, we describe the critical behavior near the quantum
transitions. Consider first, the initial onset of ${\cal
C}$-breaking from the $d$-wave superconductor (Fig.~\ref{fig1}).
For large $N$, this transition, if second-order, occurs at the
wavevector $(K=1/2,0)$ which does not equal the separation between
any two gapless Fermi points; so the charge order parameter does
not couple efficiently to the fermionic excitations. In this case,
the effective quantum critical theory contains only two real
scalars ($\phi_x$, $\phi_y$), describing the ordering along the
two axes, and has a three spacetime dimensional,
`relativistically' invariant action with the symmetry of the $Z_4$
clock model. For smaller $N$, we consider it likely that the
initial ${\cal C}$ breaking will occur at a wavevector $(K,0)$,
which is incommensurate with the underlying lattice, but which
does exactly equal the separation between gapless Fermi points in
the superconductor. The critical quantum field theory will now
contain two complex scalars (($\Phi_x$, $\Phi_y$)--their phases
represent the ability to freely slide the charge-density wave with
respect to the lattice) coupled to the four `Dirac' fermions of
the $d$-wave superconductor. Its effective action has the form $S
= S_f + S_{\Phi} + S_{\lambda}$; $S_f$ is the fermion bilinear of
the $d$-wave superconductor containing gapless Fermi points at
$(K/2,K/2)$, $(-K/2,K/2)$, $(-K/2,-K/2)$ and $(K/2,-K/2)$, and we
will denote the components of $c_{i\alpha}$ in the vicinity of
these points by $f_{a\alpha}$ respectively ($a=1\ldots 4$);
$S_{\Phi}$ contains second order spatial and time derivatives of
$\Phi_{x,y}$ and polynomial interaction terms, all invariant under
the uniform phase change $\Phi_{x,y} \rightarrow e^{i\theta_{x,y}}
\Phi_{x,y}$ and under $\Phi_x \leftrightarrow \Phi_y$;
$S_{\lambda}$ couples the $f_{a \alpha}$ and $\Phi_{x,y}$, and the
symmetries allow the following two independent terms, free of
gradients:
\begin{eqnarray}
&& \lambda_1 {\cal J}^{\alpha\beta} \left(
\Phi_x f_{1\alpha} f_{4 \beta}
+ \Phi_x^{\ast} f_{2 \alpha} f_{3 \beta}
+ \Phi_y f_{2\alpha} f_{1 \beta}
+ \Phi_y^{\ast} f_{3 \alpha} f_{4 \beta} \right) \nonumber \\
&& + \lambda_2 \left( \Phi_x f_{2\alpha}^{\dagger} f_{1 \alpha}
+ \Phi_x f_{3\alpha}^{\dagger} f_{4 \alpha}
+ \Phi_y f_{4 \alpha}^{\dagger} f_{1 \alpha}
+ \Phi_y f_{3 \alpha}^{\dagger} f_{2 \alpha}\right)
\nonumber
\end{eqnarray}
and their Hermitian conjugates. We propose that it is this quantum
field theory, describing the $T=0$ transition at which $\langle
\Phi_{x,y} \rangle$ become non-zero in the presence of
superconductivity, whose $T>0$ correlators describe the observed
quantum-critical scaling of the fermion momentum distribution
function~\cite{valla}. Direct observation of charge fluctuations
at wavevectors $(K,0)$, $(0,K)$, with $K$ consistent with
photoemission, will be a test of this scenario.

We turn to the quantum transition where ${\cal M}$ symmetry is broken
which is located at lower $\delta$. Assuming that this
is at a point where the fermion spectrum is already fully
gapped, or the separation between any gapless Fermi points is not
equal to the spin ordering wavevector, we can conclude that this
transition is described by the
relativistic quantum $O(3)$ non-linear sigma
model (for $N=1$). Such a scenario provides a natural
explanation for the crossovers in NMR experiments~\cite{imai}.

This paper has used quantitative calculations of a microscopic model
in a large-$N$ limit
to motivate a scenario in which superconducting, spin- and
charge-density wave instabilities compete as the system evolves
from an insulating antiferromagnet to a $d$-wave superconductor.
Many aspects are consistent with recent experiments, and more
stringent tests should be possible in the future.

We thank V.~Emery, M.~P.~A.~Fisher, D.~Scalapino and S.~White for
useful discussions. This research was supported by US NSF Grant No
DMR 96--23181 and by the DFG (VO 794/1-1).

\end{document}